\documentclass[11pt, a4paper]{article}

\usepackage{amsmath, bm, amssymb, mathrsfs}
\usepackage{dsfont}
\usepackage{cite}
\usepackage{braket}

\usepackage[top=25mm, bottom=25mm, left=25mm, right=25mm]{geometry}
\usepackage{titlesec}
\titlespacing*{\section}{0pt}{1ex}{-0.1ex}
\titlespacing*{\subsection}{0pt}{1ex}{-0.1ex}
\titlespacing*{\subsubsection}{0pt}{0ex}{-0.2ex}

\usepackage[colorlinks=true, linkcolor=blue, urlcolor=blue, citecolor=blue]{hyperref}

\usepackage{tikz}
\usetikzlibrary{arrows.meta}
\usetikzlibrary{decorations.pathmorphing}

\begin{document}
\thispagestyle{empty}
\begin{flushright}
\tt CP3-18-68
\end{flushright}

\vskip 40pt
\begin{center}

{\LARGE\sffamily\bfseries Gravitational Scattering in the High\raisebox{1pt}{-}Energy Limit}
\vskip 40pt

{\large\sc Zhengwen Liu}
\\
\vskip 10pt
{\it Center for Cosmology, Particle Physics and Phenomenology (CP3), \\
UCLouvain, 1348 Louvain-la-Neuve, Belgium}
\\
\vskip 8pt
{\it Email:~}\href{mailto:zhengwen.liu@uclouvain.be}{\tt zhengwen.liu@uclouvain.be}
\end{center}

\vskip 20mm
\begin{quote}
Any gravitational scattering amplitude takes a remarkably simple factorized form at tree level in multi-Regge kinematics (MRK), where the produced particles are strongly ordered in rapidity. Very recently, it was shown that the scattering equations also have a very simple structure in MRK. In this paper we study Einstein gravity amplitudes in MRK in the framework of the scattering equations. We present a derivation of the multi-Regge factorization of tree-level amplitudes with any number of external gravitons and any helicity configuration.

\vskip 90mm
\noindent{\sc\today}
\end{quote}

\newpage

\section{Introduction}

Inspired by Witten's twistor string theory \cite{Witten:2003nn}, enormous advances have been made in understanding the mathematical structures behind scattering amplitudes.
Among them, the tree-level S-matrix in many theories is expressed as a multiple contour integral in a large class of new formulations, e.g.\,\cite{Roiban:2004yf,Cachazo:2012da,Cachazo:2012kg,Cachazo:2012pz,Skinner:2013xp} (see also \cite{Atiyah:2017erd} for a recent review).
Along this line, recent progress was achieved by Cachazo, He and Yuan (CHY) who proposed a new framework where any tree-level scattering amplitudes in massless field theories in arbitrary spacetime dimensions is reformulated as a multiple integral over the moduli space of Riemann spheres with $n$ marked points \cite{Cachazo:2013iaa,Cachazo:2013gna,Cachazo:2013hca,Cachazo:2013iea}.
The integral is completely localized on the zeroes of the {\it scattering equations} that are independent of the theory.
Various worldsheet models were constructed based on ambitwistor strings \cite{Mason:2013sva,Geyer:2014fka,Casali:2015vta,Chandia:2015xfa} that provide in particular an approach to extend the CHY formalism to loop level \cite{Geyer:2015bja,Geyer:2015jch,Geyer:2016wjx,Casali:2017mss,Geyer:2016nsh}.

As the backbone of the CHY formalism, the scattering equations are universal for all theories and play a fundamental role in this content.
It is thus important to reveal mathematical structures behind the scattering equations.
In the single-soft limit, expanding the scattering equations as well as other ingredients of CHY formulas around the soft momentum provides a beautiful framework to produce various soft theorems in many theories, for example the soft graviton theorem up to sub-sub-leading order \cite{Cachazo:2013hca, Schwab:2014xua,Afkhami-Jeddi:2014fia,Kalousios:2014uva,Zlotnikov:2014sva}.
More interestingly, based on the special behavior of the solutions of the scattering equations in the double-soft limit, many new universal double-soft theorems were obtained \cite{Cachazo:2015ksa,He:2016vfi,Saha:2017yqi}.
Similarly, the collinear limits of the amplitudes were also investigated up to sub-leading order in Yang-Mills, gravity and cubic scalar theories \cite{Nandan:2016ohb}.
It was found that the solutions of the scattering equations can be interpreted as the zeros of the Jacobi polynomials in a two-parameter family of kinematics \cite{Kalousios:2013eca}.
In the so-called ``positive region'' of the space of kinematic invariants, the scattering equations can be interpreted as the equilibrium equations for a stable system of $n{-}3$ particles on a finite real interval \cite{Cachazo:2016ror}.

Very recently, an investigation of the high-energy limit of the scattering equations has been initiated \cite{Duhr:2018ppq}.
It was observed that in multi-Regge kinematics (MRK) where the final state particles are strongly ordered in rapidity, each solution of the scattering equations displays the same hierarchy as the rapidity ordering.
More remarkably, in four dimensions in the spinor-helicity formalism, the scattering equations can be exactly solved at the leading order in the multi-Regge limit for each ``helicity configuration'' for any number of external particles.
As a result, multi-Regge factorization of gluon amplitudes was exactly derived from the scattering equations.

It is natural to expect a similar simplification to appear in gravity in the multi-Regge limit.
Indeed, a compact formula for $n$ graviton scattering at tree level in MRK was obtained from $t$-channel unitarity methods by Lipatov more than three decades ago \cite{Lipatov:1982it,Lipatov:1982vv}.
This paper aims at extending the analysis in \cite{Duhr:2018ppq} from gauge theory amplitudes to Einstein gravity theory with a goal to provide a new alternative understanding of the high-energy limit of gravitational scattering.
We first translate the Lipatov formula in the spinor-helicity language, and then derive it using the four-dimensional scattering equations.
While the framework presented in \cite{Duhr:2018ppq} is applicable to graviton amplitudes because of the universality of the scattering equations, this is highly non-trivial since graviton amplitudes have a rather complicated structure even in the MHV sector.

This paper is organized as follows.
In section \ref{review} we briefly review the multi-Regge factorization of tree-level graviton amplitudes and the scattering equation formalism, which are the two most important ingredients of this paper.
In section \ref{GR-MRK} we derive the factorized form of graviton amplitudes in the multi-Regge regime.
We start by studying amplitudes in the MHV sector in section \ref{GR-MRK-MHV}, and then extend the analysis to all helicity sectors in section  \ref{GR-MRK-NkMHV}.
We include two appendices with technical proofs omitted throughout the main text.

\section{Preliminaries}\label{review}

The goal of this paper is to perform an investigation of the multi-Regge behavior of tree-level scattering amplitudes in Einstein gravity in the framework of the scattering equations.
Therefore, we provide a brief review of multi-Regge kinematics and the scattering equations in this section before presenting the main result in subsequent sections.
We follow the notations of ref.\cite{Duhr:2018ppq} in this paper.

\subsection{Multi-Regge Kinematics}\label{review-MRK}
For a $2 \to (n{-}2)$ scattering, multi-Regge kinematics is defined as the regime where the final-state particles are strongly ordered in rapidity while having comparable transverse momenta, i.e.,
\begin{align}\label{MRK-rapidity}
  y_3 \gg y_4 \gg \cdots \gg y_n \quad\text{and}\quad |{k_3^\perp}| \simeq|{k_4^\perp}|\simeq\ldots\simeq |{k_n^\perp}|,
\end{align}
where $k_a^\perp$ denote the transverse momenta, and in four dimensions we define the complexified transverse momenta as $k_a^{\perp}=k_a^x + i k_a^y$.
Employing lightcone coordinates $k_a=(k_a^+, k_a^-, k_a^\perp)$ with $k_a^{\pm}=k_a^0\pm k_a^z$, the strong ordering in rapidity is equivalent to a strong ordering in ${k^+}$-components as follows:
\begin{align}\label{MRK-motenta-order}
  k_3^+ \gg k_4^+ \gg \cdots \gg k_n^+.
\end{align}
It is convenient to work in the center-of-momentum frame where two incoming particles are back-to-back on the $z$-axis,
\begin{align}
  k_1 \,=\, (0, -{\kappa}; {0}), \quad k_2 \,=\, (-{\kappa}, 0; {0})\,,\quad \kappa\equiv \sqrt{s},
\end{align}
where $s$ is the square of the center-of-mass energy, and we take a convention of considering all momenta outgoing.

\begin{figure}[t]
  \centering
  \begin{tikzpicture}[scale=1]
  
  \fill[blue,opacity=0.25] (0,0) circle (0.25);
  \draw[line width=1.2pt] (0,0) circle (0.25);
  
  \draw[decorate, decoration={coil, aspect=0}, line width=1.1pt] (170:0.25) -- (170:2.55) node[left] {\Large $2$};
  \draw[decorate, decoration={coil, aspect=0}, line width=1.1pt] (10:0.25) -- (10:2.55) node[right] {\Large $3$};

  \fill[yshift=-1.7cm, green, opacity=0.20] (0,0) circle (0.25);
  \draw[yshift=-1.7cm, line width=1.2pt] (0,0) circle (0.25);
  
  \draw[yshift=-1.7cm,decorate, decoration={coil,aspect=0}, line width=1.1pt] (0:0.25) -- (0:2.55) node[right] {\Large $4$};

  \draw[yshift=-0.25cm, decorate,decoration=zigzag, line width=1.1pt] (0,0) -- (0,-1.2);
  
  
   \draw[yshift=-1.95cm, decorate,decoration=zigzag, line width=1.1pt] (0,0) -- (0,-0.75);
   
   \draw[yshift=-2.77cm, dashed, line width=1.1pt] (0,0) -- (0,-0.75);
   
   \draw[yshift=-4.25cm, decorate,decoration=zigzag, line width=1.1pt] (0,0) -- (0,0.75);
   
  \fill[yshift=-4.5cm, green, opacity=0.20] (0,0) circle (0.25);
  \draw[yshift=-4.5cm, line width=1.2pt] (0,0) circle (0.25);
  
  \draw[yshift=-4.5cm,decorate, decoration={coil,aspect=0}, line width=1.1pt] (0:0.25) -- (0:2.55) node[right] {\Large $n{-}1$};
  \draw[yshift=-4.75cm, decorate,decoration=zigzag, line width=1.1pt] (0,0) -- (0,-1.2);
  
  \fill[yshift=-6.2cm, blue,opacity=0.25] (0,0) circle (0.25);
  \draw[yshift=-6.2cm, line width=1.2pt] (0,0) circle (0.25);
  
  \draw[yshift=-6.2cm,decorate, decoration={coil,aspect=0}, line width=1.1pt] (190:0.25) -- (190:2.55) node[left] {\Large $1$};
  \draw[yshift=-6.2cm,decorate, decoration={coil,aspect=0}, line width=1.1pt] (-10:0.25) -- (-10:2.55) node[right] {\Large $n$};
  
\end{tikzpicture}
\caption{\label{Fig-mrk-gr}The factorized form of a tree-level amplitude of gravitons in multi-Regge kinematics.}
\end{figure}
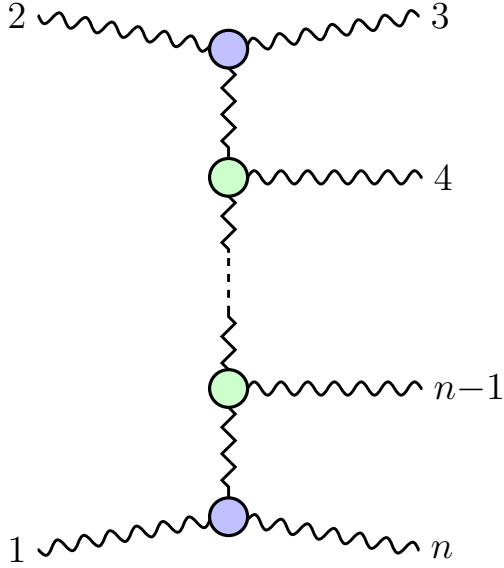

In this kinematical regime, the tree-level scattering amplitude in gravity takes a surprisingly simple factorized form: any $n$ graviton amplitude is given by only one Feynman graph with two kinds of effective vertices (see figure~\ref{Fig-mrk-gr}) \cite{Lipatov:1982it,Lipatov:1982vv}.
More precisely, one has
\begin{align}\label{MRK-GR-main}
  {\cal M}_n \,&\simeq\, -s^2\, {\cal C}(2;3) {-1 \over |{q_{4}^\perp}|^2} {\cal V}(q_{4}; 4; q_{5}) \cdots
  {-1 \over |{q_{n-1}^\perp}|^2} {\cal V}(q_{n-1}; n{-}1; q_{n}) {-1 \over |{q_{n}^\perp}|^2}\, {\cal C}(1; n)\,,
\end{align}
where we define $q_{a} = \sum_{i=2}^{a-1} k_i$ with $4\le a \le n$.
Here some overall factor including the Gravitational constant $\kappa^2=8\pi G_N$ has been stripped off.
The effective graviton-graviton-Reggeon\footnote{In this paper, Reggeon denotes `Reggeized graviton'.} vertex reads
\begin{align}\label{vertex-GGR}
  \Gamma_{\mu\nu,\alpha\beta} \,\equiv\,  \Gamma_{\mu\alpha}\Gamma_{\nu\beta} + \Gamma_{\mu\beta}\Gamma_{\nu\alpha},
\end{align}
which is manifestly a double copy of the gluon-gluon-(Reggeized gluon) vertex defined as
\begin{align}
  \Gamma_{23}^{\mu\alpha} \,&=\, - \eta^{\mu\alpha} + {2(k_1^\mu k_2^\alpha - k_3^\mu k_1^\alpha)  \over s}
  + s_{23}\, {2k_1^\alpha k_1^\mu \over s^2},
  \\
  \Gamma_{1n}^{\mu\alpha} \,&=\, - \eta^{\mu\alpha} + {2(k_2^\mu k_1^\alpha - k_n^\mu k_2^\alpha)  \over s}
  + s_{1n}\, {2k_2^\alpha k_2^\mu \over s^2}.
\end{align}
Similarly, the effective Reggeon-Reggeon-graviton vertex can also be obtained as the double copy of gauge theory vertices:
\begin{align}\label{vertex-RGR}
  \Gamma_{i}^{\mu\nu}(q_i,q_{i+1}) \,&\equiv\, 2\big(C_i^\mu C_i^\nu - N_i^\mu N_i^\nu\big),
\end{align}
where $C^\mu$ is the famous Lipatov vertex of (Reggeized gluon)-(Reggeized gluon)-gluon in QCD \cite{Lipatov:1976zz}
\begin{align}
  C_i^\mu(q_i,q_{i+1}) \,=\, - (q_i^\perp)^\mu -  (q_{i+1}^\perp)^\mu
  + \left( {2q_{i+1}^2 \over s_{1i}} - {s_{2i} \over s} \right) k_1^\mu
  - \left( {2q_i^2 \over s_{2i}} + {s_{1i} \over s} \right) k_2^\mu
\end{align}
with $(q^\perp)^\mu \equiv (0, 0; q^\perp)$, while $N^\mu$ is the so-called QED Bremsstrahlung vertex:
\begin{align}
  N_i^\mu(q_i,q_{i+1}) \,=\, \sqrt{q_i^2 q_{i+1}^2} \, \left( {k_1^\mu \over s_{1i}} - {k_2^\mu \over s_{2i}} \right).
\end{align}

The contractions between vertices and the polarization tensors of external gravitons give the {\it gravitational impact factor} and {\it gravitational Lipatov vertex} appearing in formula \eqref{MRK-GR-main}, i.e.,
\begin{align}\label{vertices-mrk-d-dim}
\begin{aligned}
  {\cal C}(2;3) \,&=\, \Gamma_{\mu\nu,\alpha\beta}(k_2, k_3)\, \epsilon_2^{\mu\nu}(k_2)\epsilon_3^{\alpha\beta}(k_3),   \\
  {\cal C}(1;n) \,&=\, \Gamma_{\mu\nu,\alpha\beta}(k_1, k_n)\, \epsilon_1^{\mu\nu}(k_1)\epsilon_n^{\alpha\beta}(k_n),   \\
  {\cal V}(q_i, i, q_{i+1}) \,&=\, \Gamma_{i, \mu\nu}(q_i, q_{i+1})\, \epsilon_i^{\mu\nu}(k_i).
\end{aligned}
\end{align}

At this point, we would like to make some comments on the effective vertices \eqref{vertex-GGR} and \eqref{vertex-RGR} in gravity in MRK.
These effective vertices have also been derived from an effective action (c.f.~e.g.~\cite{Lipatov:1991nf,Lipatov:2011ab,Lipatov:2014cka}).
It is extremely remarkable that the double copy relation between gravity and gauge theories was uncovered for the first time in MRK.
In general kinematics, Kawai, Lewellen and Tye (KLT) found that a closed string amplitude can be expressed in terms of sums of products of two open string amplitudes \cite{Kawai:1985xq}. In the field theory limit, the KLT relation naturally implies the double copy relation between amplitudes in gravity and Yang-Mills.

In the following we would like to translate the Lipatov formula \eqref{MRK-GR-main} to modern language, say spinor-helicity variables.
In four dimensions, any massless momentum can be written as a product of two spinors with opposite chirality, i.e.\,$k_i^{\alpha\dot\alpha}=\lambda_i^\alpha\tilde\lambda_i^{\dot\alpha}$.
To be explicit, in this paper two-component spinors are defined as
\begin{align}\label{spinor-def}
\begin{aligned}
  \lambda_1 \,=\, - \tilde\lambda_1 \,=\, 
  \left(\begin{matrix}  0  \\  \sqrt{\kappa} \end{matrix}\right),
  &\quad
  \lambda_2 \,=\, -\tilde\lambda_2 \,=\, 
  \left(\begin{matrix}  \sqrt{\kappa}  \\  0  \end{matrix}\right),
  \\
  \lambda_a \,=\, {1 \over \sqrt{k_a^+}}
  \left(\begin{array}{l} {k_a^+}  \\  k_a^\perp  \end{array}\right),
  &\quad
  \tilde\lambda_a \,=\, {1 \over \sqrt{k_a^+}}
  \left(\begin{array}{l} {k_a^+}  \\  {k_a^{\perp^\ast}}  \end{array}\right),
  \quad 3\le a\le n.
\end{aligned}
\end{align}
Similarly, we can also write the polarization vectors in terms of spinor-helicity variables as follows:
\begin{align}\label{polarization-spinor}
  \epsilon_i^{+,\alpha\dot\alpha}  \,=\, {\lambda_x^\alpha \tilde\lambda_i^{\dot\alpha} \over \braket{x\,i}}, \qquad
  \epsilon_i^{-,\alpha\dot\alpha}  \,=\,  {\lambda_i^\alpha \tilde\lambda_x^{\dot\alpha} \over [{i\,x}]},
\end{align}
where it is free to choose the reference spinor $\lambda_x$ or $\tilde\lambda_x$.
A polarization tensor can be expressed in terms of the symmetric-traceless tensor product of two polarization vectors.
Therefore we can write the graviton polarization tensor in spinor variables as\footnote{This representation for graviton polarization tensors satisfies $\epsilon_i^+ \cdot \epsilon_i^- \equiv \epsilon_i^{+,\mu\nu} \epsilon^-_{i,\mu\nu} = 1$.
}
\begin{align}
  \epsilon_i^{+,\alpha\dot\alpha\beta\dot\beta}  \,=\, 
  {\lambda_x^\alpha \tilde\lambda_i^{\dot\alpha} \over \braket{x\,i}} {\lambda_y^\beta \tilde\lambda_i^{\dot\beta} \over \braket{y\,i}}
  + (x \leftrightarrow y),
  \\
  \epsilon_i^{-,\alpha\dot\alpha\beta\dot\beta}  \,=\,  
  {\lambda_i^\alpha \tilde\lambda_x^{\dot\alpha} \over [{i\,x}]}{\lambda_i^\beta \tilde\lambda_y^{\dot\beta} \over [{i\,y}]}
  + (x \leftrightarrow y),
\end{align}
where $x$ and $y$ stand for arbitrary reference spinors.

Using the spinor-helicity variables defined above, it is easy to compute the gravitational impact factors and Lipatov vertices defined in \eqref{vertices-mrk-d-dim}.
A straightforward calculation gives
\begin{align}\label{gr-impact-factor}
\begin{aligned}
  {\cal C}(2^+;3^+) \,=\, {\cal C}(2^-;3^-) \,=\, 0,\quad
  {\cal C}(2^-;3^+)  \,&=\, {\cal C}(2^+;3^-) \,=\,1,\\
  {\cal C}(1^+;n^+) \,=\, {\cal C}(1^-;n^-) \,=\, 0, \quad
  {\cal C}(1^-;n^+)  \,&=\, {\cal C}(1^+;n^-)^\ast \,=\, \left({k_n^{\perp^\ast} \over k_n^\perp}\right)^2,
\end{aligned}
\end{align}
and
\begin{align}
  {\cal V}\big( q_{a}; a^+; q_{a+1} \big) 
  \,=\, {\cal V}\big( q_{a}; a^-; q_{a+1} \big)^\ast 
  \,&=\, {q_a^{\perp^\ast} \big( q_{a}^{\perp^\ast} q_{a+1}^{\perp} - q_{a}^{\perp} q_{a+1}^{\perp^\ast} \big) q_{a+1}^{\perp} \over (k_a^\perp)^2}
  \nonumber
  \\
  \,&=\, {q_a^{\perp^\ast} \big( k_{a}^{\perp} q_{a}^{\perp^\ast}  - k_{a}^{\perp^\ast} q_{a}^{\perp} \big) q_{a+1}^{\perp} \over (k_a^\perp)^2}.
\end{align}
We see from \eqref{gr-impact-factor} that helicity is conserved by the impact factors, like in gauge theory.

\subsection{Scattering equation formalism}\label{review-SE}

The aim of this paper is to show that the multi-Regge limit of graviton amplitudes summarized in the previous section can be elegantly derived from the scattering equations.
Let us continue to review the scattering equation formalism in the following.
The cornerstone is of course the {\it scattering equations} \cite{Fairlie-Roberts-1972, Gross-Mende-1987, Witten:2004cp, Cachazo:2013iaa, Fairlie:2008dg, Cachazo:2013gna}
\begin{align}\label{SE-chy}
  f_a \,=\, \sum_{b\neq a} {k_a\cdot k_b \over \sigma_{a}  {-} \sigma_{b}} \,=\, 0\,,
  \qquad a\in\{1, 2, \ldots, n\},
\end{align}
where $\sigma_a\in \mathbb{CP}^1$ denote the positions of the punctures in the moduli space $\mathfrak{M}_{0,n}$ of Riemann spheres with $n$ marked points.
The scattering equations are invariant under M\"{o}bius transformations, and only $n{-}3$ of the $n$ equations are independent.

In four dimensions, the use of spinor-helicity variables usually leads to significant simplifications.
Indeed, it has been shown that in four dimensions the scattering equations can be decomposed into different `helicity sectors'.
One of constructions is \cite{Geyer:2014fka}
\begin{align}\label{SE-4d-twistor}
\begin{aligned}
  \bar{\cal E}^{\dot\alpha}_I \,&=\,  \tilde\lambda_I^{\dot\alpha}  - \sum_{i\in\mathfrak{P}} {t_I t_i \over \sigma_{I} {-} \sigma_{i}}\tilde\lambda_i^{\dot\alpha} \,=\, 0,
  ~~I\in\mathfrak{N}\,,\\
  {\cal E}^{\alpha}_i \,&=\,  \lambda_i^{\alpha} - \sum_{I\in\mathfrak{N}} {t_i t_I \over \sigma_{i} {-} \sigma_{I}}\lambda^\alpha_I \,=\, 0,
  ~~i\in\mathfrak{P},
\end{aligned}
\end{align}
where $\mathfrak{N}$ is a subset of $\{1,\ldots,n\}$ with length $k\in\{2,\ldots,n{-}2\}$ and $\mathfrak{P}$ is the corresponding complement.
We refer to equations \eqref{SE-4d-twistor} as the {\it four-dimensional scattering equations} of sector $k$; an important property is that only the equations in the sector $k$ are needed for N$^{k-2}$MHV amplitudes.
Similarly, the system of the four-dimensional scattering equations owns a $\text{GL}(2,{\mathbb C}) = \text{SL}(2,{\mathbb C}) \times \text{GL}(1,{\mathbb C})$ symmetry.

As shown in the previous section, the multi-Regge limit is most naturally defined in terms of lightcone variables.
It is thus natural to write the scattering equations in terms of lightcone coordinates.
While one can get equations with lightcone coordinates by simply substituting \eqref{spinor-def} into the spinor-valued scattering equations \eqref{SE-4d-twistor}, we show that one can obtain a nicer form by fixing the redundancy and rescaling variables and equations. 
First, let us use the GL$(2, \mathbb{C})$ redundancy to fix four variables as follows:
\begin{align}\label{gauge-fixing}
  \sigma_1 \,=\,  0, \qquad  \sigma_2 \,=\, t_2 \,\to\, \infty, \qquad
  t_1 \,=\, -1.
\end{align}
Here we always use the convention where $\{1, 2\} \subseteq \mathfrak{N}$, and we define $\overline{\mathfrak{N}} \equiv \mathfrak{N}\backslash\{1,2\}$.
Moreover, we follow the convention that elements of $\mathfrak{P}$ and $\overline{\mathfrak{N}}$ are denoted by small and capital letters respectively, e.g.~$i\in \mathfrak{P}$ and $I\in \overline{\mathfrak{N}}$.
Second, we perform a rescaling for the $t_a$ variables as follows:
\begin{align}\label{t-resacling}
  t_i \,=\, \tau_i\sqrt{k_i^+ \over {\kappa}}, \qquad 
  t_I \,=\, \tau_I\, {\sqrt{\kappa\,k_I^+}  \over k_I^\perp}.
\end{align}
Then let us also perform a rescaling for the scattering equations according to:
\begin{align}
\begin{aligned}
  {\cal S}_i^{1} \,\equiv\,  {1 \over \lambda^1_i}\, {\cal E}^{1}_i, \quad
  {\cal S}_i^{2} \,\equiv\,  {\lambda^1_i \over k_i^\perp}\, {\cal E}^{2}_i; \qquad
  \bar{\cal S}_I^{\dot 1} \,&\equiv\,  \lambda^2_I\,\bar{\cal E}^{\dot 1}_I,  \quad
  \bar{\cal S}_I^{\dot 2} \,\equiv\,  \lambda^1_I\,\bar{\cal E}^{\dot 2}_I;
  \\
  \bar{\cal S}_1^{\dot 1}  \,\equiv\,  \lambda^2_1\,\bar{\cal E}^{\dot 1}_1, \quad
  \bar{\cal S}_1^{\dot 2} \,\equiv\,  \lambda^2_1\,\bar{\cal E}^{\dot 2}_1; \qquad
  \bar{\cal S}_2^{\dot 1} \,&\equiv\,  \lambda^1_2\,\bar{\cal E}^{\dot 1}_2, \quad
  \bar{\cal S}_2^{\dot 2} \,\equiv\,  \lambda^1_2\,\bar{\cal E}^{\dot 2}_2.
\end{aligned}
\end{align}
As a consequence, we obtain a set of equations that contain only the terms linear in $k_a^+$.
Explicitly, we have
\begin{align}\label{SE-lightcone}
\begin{aligned}
  {\cal S}_i^{1}  \,&=\,
  1  + \tau_i  - \sum_{I\in\overline{\mathfrak{N}}} {\tau_i \tau_I \over \sigma_i {-} \sigma_I} {k_I^+  \over  k_I^\perp}
  \,=\, 0,
  \\
  \bar{\cal S}_I^{\dot 1}  \,&=\,
  {k_I^\perp}  - \sum_{i\in\mathfrak{P}} {\tau_i \tau_I \over \sigma_I {-} \sigma_i} {k_i^+}
  \,=\, 0,
  \\
  \bar{\cal S}_1^{\dot 1} \,&=\, - \sum_{i\in\mathfrak{P}} {\tau_i \over \sigma_i} {k_i^+}
  \,=\, 0,
  \\
  \bar{\cal S}_2^{\dot 1}  \,&=\, -{\kappa}  - \sum_{i\in\mathfrak{P}} \tau_i {k_i^+}
  \,=\, 0,
\end{aligned}
\qquad
\begin{aligned}
  {\cal S}_i^{2}  \,&=\,
  1 + {k_i^+ \over k_i^\perp}{\tau_i \over \sigma_i}
  - {k_i^+ \over k_i^\perp} \sum_{I\in\overline{\mathfrak{N}}} {\tau_i \tau_I \over \sigma_i {-} \sigma_I}  
  \,=\, 0;
  \\
  \bar{\cal S}_I^{\dot 2}  \,&=\,
  k_I^{\perp^\ast}  - {k_I^+ \over k_I^\perp} \sum_{i\in\mathfrak{P}} {\tau_i \tau_I \over \sigma_I {-} \sigma_i}  k_i^{\perp^\ast}
  \,=\, 0;
  \\
  \bar{\cal S}_1^{\dot 2} \,&=\, - {\kappa}  - \sum_{i\in\mathfrak{P}} {\tau_i \over \sigma_i} k_i^{\perp^\ast}
  \,=\, 0;
  \\
  \bar{\cal S}_2^{\dot 2} \,&=\, - \sum_{i\in\mathfrak{P}} \tau_i k_i^{\perp^\ast}
   \,=\, 0.
\end{aligned}
\end{align}
We would like to emphasize that no limit has been applied to these equations, and they are completely equivalent to the four-dimensional scattering equations in \eqref{SE-4d-twistor}, up to fixing the GL$(2,\mathbb{C})$ redundancy according to \eqref{gauge-fixing} and performing the rescaling in \eqref{t-resacling}. 

In terms of lightcone variables, the formula for tree-level N$^{k-2}$MHV graviton amplitudes reads \cite{Geyer:2014fka}
\begin{align}\label{GR-amp-lc}
  {\cal M}_{n,k}
  \,&=\, - s^2\,
  \Bigg(\int \prod_{a=3}^{n} {d\tau_a d \sigma_a \over \tau_a^3} \Bigg)\,
  \Bigg(\prod_{I\in\overline{\mathfrak{N}}} {(k_I^\perp)^3 \over k_I^+} \delta^{2}\big( \bar{\cal S}^{\dot\alpha}_I \big) \Bigg)
  \Bigg(\prod_{i\in\mathfrak{P}} {1 \over k_i^+ k_i^\perp} \delta^2\big( {\cal S}^\alpha_i \big)\Bigg)
  {\cal I}^\text{(GR)}_{n,k}.
\end{align}
If we assume that gravitons 1 and 2 carry negative helicity and use $\mathfrak{N}$ and $\mathfrak{P}$ to collect the babels of negative and positive helicity gravitons, the integrand function in formula \eqref{GR-amp-lc} takes the following simple form:
\begin{align}\label{}
  {\cal I}^\text{(GR)}_{n,k} \,=\, {\det}'\,\overline{\mathsf H}\,{\det}'\,{\mathsf H},
\end{align}
where $\det'$ denotes the minor with any one column and row deleted, and ${\mathsf H}$ is the symmetric $k\times k$ matrix defined as
\begin{align}\label{}
  {\mathsf H}_{12} &= -1, ~~
  {\mathsf H}_{1I} = -{\tau_I \over \sigma_I}{k_I^+ \over k_I^\perp}, ~~
  {\mathsf H}_{2I} = \tau_I, ~~
  {\mathsf H}_{IJ} = {\tau_I\tau_J \over \sigma_I {-} \sigma_J}\left( {k_I^+ \over k_I^\perp} - {k_J^+ \over k_J^\perp} \right)
  ~\text{for}~I\ne J,
  \\
  {\mathsf H}_{aa} &= -\sum_{b\in\mathfrak{N},b\ne a} {\mathsf H}_{ab}, ~~a\in\mathfrak{N},
\end{align}
and the $\overline{\mathsf H}$ is the symmetric $(n{-}k)\times (n{-}k)$ matrix:
\begin{align}\label{}
  \overline{\mathsf H}_{ij} \,&=\, {\tau_i\tau_j \over \sigma_i {-} \sigma_j} \Big( {k_i^+}{k_j^{\perp^\ast}} - {k_j^+}{k_i^{\perp^\ast}} \Big)
  ~\text{for}~i\ne j,
  \\
   \overline{\mathsf H}_{ii} \,&=\, -\sum_{j\in\mathfrak{P},j\ne i} \overline{\mathsf H}_{ij}.
\end{align}
In formula \eqref{GR-amp-lc}, we have eliminated four of the scattering equations and identify them with the momentum conservation delta-functions, i.e.,
\begin{align}\label{SEs-12-momentum-conserv}
  \delta^2\big( \bar{\cal S}^{\dot\alpha}_1 \big)  \delta^2\big( \bar{\cal S}^{\dot\alpha}_2 \big)
  \,=\, \delta^4\left(\sum_{a=1}^{n} k_a^\mu \right).
\end{align}

\section{Gravitational scattering in MRK}\label{GR-MRK}
The goal of this section is to present an alternative derivation of the multi-Regge factorization of graviton amplitudes based on the scattering equations.

It was conjectured in \cite{Duhr:2018ppq} that in MRK all solutions of the scattering equations satisfy
\begin{align}\label{conjecture}
  {1 \over \sigma_a {-} \sigma_b} \,\simeq\, {1 \over \sigma_a}~~\text{when}~~a<b, \quad
  \tau_a \,=\, {\cal O}(k_a^\perp).
\end{align}
Expanding the formula for graviton amplitudes \eqref{GR-amp-lc} to leading power in MRK according to \eqref{conjecture}, it gets significantly simplified
\begin{align}\label{GR-amp-MRK}
  {\cal M}_{n,k}
  \,&\simeq\, - s^2\,
  \Bigg(\int \prod_{a=3}^{n} {d\tau_a d \zeta_a \over \tau_a^2\zeta_a^2} \Bigg)\,
  \Bigg(\prod_{I\in\overline{\mathfrak{N}}} (k_I^\perp)^2\, \delta^{2}\big( \bar{\cal S}^{\dot\alpha}_I \big) \Bigg)
  \Bigg(\prod_{i\in\mathfrak{P}} {1 \over (k_i^\perp)^2} \delta^2\big( {\cal S}^\alpha_i \big)\Bigg)
  {\det}'\,\overline{\mathsf H}\,{\det}'\,{\mathsf H},
\end{align}
where the new variables are defined as
\begin{align}\label{zeta-vars}
  \zeta_a \,&\equiv {k_a^+ \over k_a^\perp}\,{\tau_a \over \sigma_a}, \quad 3 \le a \le n\,.
\end{align}
Here we fix $\{3, n\} \subseteq \mathfrak{P}$ as a convention.
The leading order approximation of the scattering equations in the multi-Regge limit is given by
\begin{align}\label{SE-MRK}
\begin{aligned}
   {\cal S}_i^{1}  \,=\, 1 + \tau_i\,\Bigg(1+ \sum_{I\in\overline{\mathfrak N}_{<i}} \zeta_I \Bigg) = 0\,,\qquad 
  & \bar{\cal S}_I^{\dot{1}} \,=\, {k_I^\perp}  + \tau_I \sum_{i\in\mathfrak{P}_{<I}} \zeta_i\, {k_i^\perp}  = 0\,,
  \\
  {\cal S}_i^{2}  \,=\, 1 + \zeta_i\,\Bigg(1 - \sum_{I\in\overline{\mathfrak N}_{>i}} \tau_I\Bigg)   = 0\,,\qquad
  & \bar{\cal S}_I^{\dot{2}} \,=\, k_I^{\perp^\ast}  - \zeta_I\,\sum_{i\in\mathfrak{P}_{>I}} \tau_i\, k_i^{\perp^\ast}  = 0.
\end{aligned}
\end{align}
They have the following unique solution:
\begin{align}\label{MRK-sol}
\begin{aligned}
  \tau_{I} \,&=\, {k_{I}^\perp  \over q_{I+1}^\perp} \prod_{J\in\mathfrak{N}_{>I}} {q_{J}^\perp  \over q_{J+1}^\perp},
  \qquad
  \zeta_{I}  \,=\, \Bigg( {k_{I}^{\perp}  \over q_{I}^{\perp}}\Bigg)^{\!\!\ast}\, \Bigg(\prod_{J\in\mathfrak{N}_{<I}} {q_{J+1}^{\perp} \over q_{J}^{\perp}}\Bigg)^{\!\!\ast},
  \\
  \tau_i  \,&=\, - \Bigg(\prod_{I\in\overline{\mathfrak{N}}_{<i}} {q_{I}^{\perp} \over q_{I+1}^{\perp}}\Bigg)^{\!\!\ast},
  \qquad
  \zeta_i  \,=\, - \prod_{I\in\overline{\mathfrak{N}}_{>i}} {q_{I+1}^\perp  \over q_{I}^\perp}.
\end{aligned}
\end{align}

In the following we show that we can obtain the multi-Regge factorization of graviton amplitudes by localizing integrals in \eqref{GR-amp-MRK} to the unique solution \eqref{MRK-sol}.

In \cite{Duhr:2018ppq}, it is shown that for any function ${\cal F}(\tau_a,\zeta_a)$ of $\tau_a$ and $\zeta_a$ we have
\begin{align}\label{theorem-MRK}
  \!\!\!\!
  \begin{aligned}
  \boxed{
  \int \prod_{a=3}^{n} {d\zeta_a d\tau_a \over \zeta_a \tau_a} \!
  \prod_{I\in\overline{\mathfrak{N}}}  \!\delta^{2}\big( \bar{\cal S}^{\dot\alpha}_I \big)
  \prod_{i\in\mathfrak{P}} \!\delta^2\big( {\cal S}^\alpha_i \big)
  {\cal F}(\zeta_a, \!\tau_a)
  = (-1)^n\!
  \left(\prod_{I\in\overline{\mathfrak{N}}} 
  {1 \over |k_I^\perp|^2}  {q_I^\perp q_{I+1}^{\perp\ast} \over  q_I^{\perp\ast} q_{I+1}^\perp}  \right)
  \!{\cal F}(\zeta_a, \!\tau_a)\Big|_\text{\eqref{MRK-sol}}.
  }
  \end{aligned}
\end{align}

For gluon amplitudes, ${\cal F}$ is just a constant factor without the dependence of $\zeta_a$ and $\tau_a$ \cite{Duhr:2018ppq}, i.e.,
\begin{align}\label{F-fun-YM}
  {\cal F}^\text{(YM)} \,=\,  -s\,
  \Bigg( \prod_{i\in\mathfrak{P}} {1 \over k_i^\perp} \Bigg)
  \Bigg( \prod_{I\in\overline{\mathfrak N}} k_I^\perp \Bigg).
\end{align}
This paper focuses on the gravitational scattering whose ${\cal F}$ function takes
\begin{align}\label{F-fun-GR}
  {\cal F}^\text{(GR)} \,=\,  -s^2\,
  \Bigg( \prod_{i\in\mathfrak{P}} {1 \over (k_i^\perp)^2}\, {1 \over \zeta_i \tau_i} \Bigg){\det}'\,\overline{\mathsf H}\,
  \Bigg( \prod_{I\in\overline{\mathfrak N}} {(k_I^\perp)^2 \over  \zeta_I \tau_I} \Bigg) {\det}'\,{\mathsf H}.
\end{align}
The main task of the rest part of the paper is to calculate this quantity on the support of the unique solution \eqref{MRK-sol} of the four-dimensional scattering equations in MRK.


\subsection{MHV sector}\label{GR-MRK-MHV}
Let us first consider the ${\cal F}$ function defined in \eqref{F-fun-GR} in the MHV sector.
The experience from the MHV sector will be useful for evaluating the determinants in other N$^k$MHV sectors.

In this case, $\mathfrak{P}=\{3,\ldots,n\}$, the solution \eqref{MRK-sol} becomes $\zeta_i=\tau_i=-1$, and we have
\begin{align}\label{}
  \overline{\mathsf H}_{ij} \,&=\,  ({k_j^\perp}\zeta_j)\,(k_i^{\perp^\ast}\tau_i)
  \,=\, (k_i^{\perp} {k_j^\perp})\,x_i ~~\text{when}~~i>j,
  \\
  \overline{\mathsf H}_{ij} \,&=\, (k_i^{\perp} {k_j^\perp})\,x_j ~~\text{when}~~ i<j,
  \\
  \overline{\mathsf H}_{ii} \,&=\, - \sum_{j\in\mathfrak{P}, j\ne i}\overline{\mathsf H}_{ij} 
  \,=\,  ({k_i^\perp})^2\, \big(x_i + v_i\big),
\end{align}
where we define
\begin{align}\label{}
  x_a \,\equiv\, {k_a^{\perp^\ast} \over k_a^\perp}, \qquad
  v_a \,&\equiv\,  {k_a^{\perp} q_a^{\perp^\ast}  \!\!- q_a^{\perp} k_a^{\perp^\ast} \over (k_a^\perp)^2}.
\end{align}

Let us choose to delete the first column and row corresponding to the particle label `3' from the matrix $\overline{\mathsf H}$. Then the reduced determinant can be written as
\begin{align}\label{det-pre-mhv}
  {\det}'\, \overline{\mathsf H} \,=\, {1 \over ({k_3^\perp})^2}\Bigg( \prod_{i\in{\mathfrak P}}  ({k_i^\perp})^2 \Bigg)
  \det \bar\phi,
\end{align}
with
\begin{align}\label{MHV-phi-matrix-MRK}
   \bar\phi \,=\, 
   \left(\begin{matrix}
       x_4 {+} v_4  &  x_5 & \cdots & x_{n-1} & x_n \\
       x_5  & x_5 {+} v_5  & \cdots & x_{n-1} & x_n \\
       \vdots  & \vdots   & \ddots & \vdots & \vdots \\
       x_{n-1}  & x_{n-1}  & \cdots  & x_{n-1} {+} v_{n-1} & x_n \\
       x_n  & x_n & \cdots & x_n & x_n
  \end{matrix}\right).
\end{align}
This matrix is nothing but the leading order approximation of Hodges' matrix\footnote{
For arbitrary external kinematics, in the MHV sector, the four-dimensional scattering equations have only one set of independent of solutions and the formula \eqref{GR-amp-lc} is simply reduced to Hodges' formula where the MHV amplitude of gravitons is given by the determinant of a symmetric matrix.}\cite{Hodges:2011wm, Hodges:2012ym} in the multi-Regge limit.
We can observe a lot of nice properties.
In particular, a conspicuous feature is that the entries $\bar\phi_{ij}$ are equal when $j<i$ for each $i$-th row.
As we now show in the following, this implies further simplification.

By performing some elementary row/column transformations of matrix, we have
\begin{align}\label{}
  \det \bar\phi \,=\, \det \bar\phi^\prime,
\end{align}
with
\begin{align}\label{}
   \bar\phi^\prime \,=\, 
   \left(\begin{matrix}
       v_4 & x_5 {-} x_4 {-} v_4 & \cdots & x_{n-1} {-} x_4 {-} v_4 & x_n {-} x_4   \\
       0 & v_5 & \cdots & x_{n-1} {-} x_5 & x_n {-} x_5   \\
       \vdots & \vdots & \ddots & \vdots & \vdots \\
       0 &  0 & \cdots & v_{n-1} & x_n {-} x_{n-1} \\
       x_n & 0 & \cdots & 0 & x_n \\
  \end{matrix}\right).
\end{align}
This is almost an upper triangular matrix.
We find it remarkable that we can nicely compute its determinant by employing the so-called the {\it matrix determinant lemma} (c.f.\,\cite{Harville-1997, Ding-Zhou-MDL-2007}, see also Appendix \ref{app-mdl} of this paper).
We first simply decompose the matrix into an upper triangular part and a matrix that has only non-zero element $x_n$ in the lower left corner.
To be more precise, we write 
\begin{align}\label{}
    \bar\phi^\prime \,=\, \varphi + \mu\,\nu^\mathrm{T}
    ~~\text{with}~~
    \mu \,=\, (0,\ldots,0, 1)^\mathrm{T},~\nu \,=\, (x_n,0,\ldots,0)^\mathrm{T},
\end{align}
where $\varphi$ is nothing but the matrix $\bar\phi'$ with replacing the first element of the last row $x_n$ by zero.
Then by making use of the matrix determinant lemma, we have
\begin{align}\label{MDL-mhv}
    \det \bar\phi' \,=\, \big(\det\varphi\big) \big(1 + \nu^\mathrm{T}\varphi^{-1}\mu\big)
    \,=\, v_4\,v_5 \cdots v_{n-1}\,x_n\,\left( 1 + x_n \big(\varphi^{-1} \big)_{1,n-3} \right).
\end{align}
In general, it seems difficult to exactly find the inverse of the matrix $\varphi$.
Fortunately, it is not hard to obtain the entry $(\varphi^{-1}) _{1,n-3}$ by induction (see Appendix \ref{app-determinents} for the details of the derivation),
\begin{align}\label{mhv-mdl-result}
  (\varphi^{-1}) _{1,n-3} \,&=\, -{k_3^\perp + k_n^\perp \over k_n^{\perp^\ast}}.
\end{align}
Plugging it into \eqref{MDL-mhv} immediately gives
\begin{align}\label{mrk-mhv-det-final}
  \det \bar\phi^\prime \,=\, -{k_3^\perp \over k_n^\perp}\, v_4\,v_5 \cdots v_{n-1}\,x_n.
\end{align}

Noting
\begin{align}\label{}
  {\cal V}(q_i, i^+, q_{i+1}) \,=\, q_{i}^{\perp^\ast} v_i\, q_{i+1}^{\perp},
  \qquad {\cal V}(1^-; n^+) \,=\, x_n^2,
\end{align}
and equations \eqref{mrk-mhv-det-final}, \eqref{det-pre-mhv}, \eqref{F-fun-GR} and \eqref{theorem-MRK}, we have
\begin{align}\label{MRK-Graviton-MHV}
  {\cal M}_n(1^-, 2^-) \,&=\, -s^2\, {\cal C}({2^-; 3^+}) \,{-1 \over |q_4^\perp|^2}\, {\cal V}(q_4, 4^+, q_{5})\,\cdots\,
  {-1 \over |q_n^\perp|^2}\, {\cal C}({1^-; n^+}).
\end{align}
We derive the correct multi-Regge factorization of any MHV amplitude.
In order to extend the analysis in the MHV sector to other helicity sectors,  at this point let us summarize some key technical ideas that have been used above.
First, we can transform the matrix into a near upper triangular form by some elementary row and column operations since the matrix has a special structure in MRK.
Then the matrix determinant lemma can be used to compute its determinant.
We show in the next section that this technique is useful for any other helicity configuration.

\subsection{All helicity configurations}\label{GR-MRK-NkMHV}

Let us first consider the positive helicity part $\mathfrak{P}$.
In MRK, the entries of $(n{-}k)\times (n{-}k)$ matrix $\overline{\mathsf H}$ take
\begin{align}\label{}
  \overline{\mathsf H}_{ij} \,&=\,  ({k_j^\perp}\zeta_j)\,(k_i^{\perp^\ast}\tau_i)
  \,=\, (k_i^{\perp}\tau_i)\, ({k_j^\perp}\zeta_j)\,x_i,
  \quad \text{for}~i>j
  \\
  \overline{\mathsf H}_{ij} \,&=\,   (k_j^{\perp}\tau_j)\, ({k_i^\perp}\zeta_i)\,x_j
  \,=\, (k_i^{\perp}\tau_i)\, ({k_j^\perp}\zeta_j)\, c_{ij},
  \quad c_{ij} \,=\, {\zeta_i \tau_j\,x_j  \over \zeta_j \tau_i},
  \quad \text{for}~i<j.
\end{align}
For diagonal elements, we have
\begin{align}\label{}
  \overline{\mathsf H}_{ii} \,&=\, - \sum_{j\in\mathfrak{P}_{<i}}\overline{\mathsf H}_{ij} - \sum_{j\in\mathfrak{P}_{>i}}\overline{\mathsf H}_{ij}
  \nonumber\\
  \,&=\, - (\tau_i  k_i^{\perp^\ast}) \sum_{j\in\mathfrak{P}, j<i} \zeta_j\,{k_j^\perp} 
  - (\zeta_i k_i^\perp) \sum_{j\in\mathfrak{P}, j>i} \tau_j \,k_j^{\perp^\ast}
  \nonumber\\
  \,&=\,- \tau_i k_i^{\perp^\ast} \sum_{j\in\mathfrak{P}, j<i} \zeta_j\,{k_j^\perp} 
  + \zeta_i {k_i^\perp} \sum_{j\in\mathfrak{P}, j \le i} \tau_j \,k_j^{\perp^\ast}
  \nonumber
  \\
  \,&=\,  ({k_i^\perp}\zeta_i)\,(k_i^\perp \tau_i)\, \big(x_i + u_i\big),
\end{align}
where one has used the momentum conservation \eqref{SEs-12-momentum-conserv} in the third line, and $u_i$ is defined as
\begin{align}\label{}
  u_i \,&\equiv\, \sum_{j\in\mathfrak{P}_{< i}} 
  \left( {\tau_j \,k_j^{\perp} \over \tau_i \,k_i^{\perp}} x_j  -  {\zeta_j\,{k_j^\perp} \over \zeta_i\,{k_i^\perp}} x_i \right).
\end{align}
Using the scattering equations \eqref{SE-MRK} and their solution \eqref{MRK-sol}, one can obtain
\begin{align}\label{}
  u_i \,&=\, v_i \,=\, {k_i^{\perp} q_i^{\perp^\ast}  - q_i^{\perp} k_i^{\perp^\ast} \over (k_i^\perp)^2}.
\end{align}

Let us choose to delete the first column and row corresponding to particle label `3', then the reduced determinant becomes nicely
\begin{align}\label{det-H-pos-x1}
  {\det}'\, \overline{\mathsf H} \,=\, \Bigg( \prod_{i\in{\mathfrak P}, i\ne 3}  ({k_i^\perp})^2 \zeta_i\,\tau_i \Bigg)
  \det \overline{\mathsf H}',
\end{align}
where
\begin{align}\label{H-matrix-mrk-pos}
   \overline{\sf H}'  \,&=\, 
   \left(\begin{matrix}
       v_{i_1} {+} x_{i_1} & c_{i_1 i_2} & c_{i_1 i_3}  & \cdots & c_{i_1 n}
       \\
       x_{i_2} & v_{i_2} {+} x_{i_2} & c_{i_2 i_3}  & \cdots & c_{i_2 n}
       \\
       x_{i_3} & x_{i_3} & v_{i_3} {+} x_{i_3}  & \cdots & c_{i_3 n}
       \\
       \vdots & \vdots & \vdots & \ddots & \vdots 
       \\
       x_{n} & x_{n} & x_{n}  & \cdots & x_{n}
       \\
   \end{matrix}\right).
\end{align}
Here labels satisfy $3<i_1<i_2<\cdots<n$.
In the case of the MHV sector, since $c_{ij}=x_j$ ($i<j$), this matrix is identical to the matrix $\bar\phi$ in \eqref{MHV-phi-matrix-MRK}.
More remarkably, they have a similar structure and share many properties.
As a consequence, we can use the same technique to calculate the determinant as in the MHV sector.
Here we summarize the result as follows (see Appendix \ref{app-determinents} for a detailed derivation):
\begin{align}\label{theorem-2}
\boxed{
  {\det}'\,\overline{\mathsf H} \,=\, \bigg(\prod_{i\in\mathfrak{P}} ({k_i^\perp})^2\, \zeta_i\, \tau_i \bigg)
  {x_n \over q_4^\perp q_n^\perp} \prod_{\substack{i\in\mathfrak{P} \\ i\ne 3, n}} v_i.
  }
\end{align}

Let us now discuss the ${k\times k}$ matrix ${\mathsf H}$.
It is reasonable to expect that the similar structure appears in this matrix such that we can compute its determinant using the matrix determinant lemma.
Let us first compute the entries of the ${\mathsf H}$ in MRK
\begin{align}\label{}
  {\mathsf H}_{12} \,&=\,  - 1,  \qquad\qquad~\,
  {\mathsf H}_{1I} \,=\, -  \zeta_I,
  \\
  {\mathsf H}_{I2} \,&=\, \big( x_I \tau_I \big) x_I^\ast, \qquad
  {\mathsf H}_{2I} \,=\, c_{2I}\, \zeta_I,  \quad  c_{2I} \,\equiv\, {\tau_I \over \zeta_I},
  \\
  {\mathsf H}_{IJ} \,&=\, \big( x_I\tau_I \zeta_J \big) x_I^\ast, \quad I>J
  \\
  {\mathsf H}_{IJ} \,&=\, c_{IJ} \big( x_I\tau_I \zeta_J \big) x_I^\ast, \quad
  c_{IJ} \,\equiv\, {\zeta_I \tau_J \over  \tau_I \zeta_J},   \quad  I<J
  \\
  {\mathsf H}_{22} \,&=\, - {\mathsf H}_{12} - \sum_{I\in\overline{\mathfrak N}} {\mathsf H}_{2I},
  \\
  {\mathsf H}_{II} \,&=\, - {\mathsf H}_{1I} - {\mathsf H}_{2I}  
  - \sum_{J\in\overline{\mathfrak N}_{<I}} {\mathsf H}_{IJ}
  - \sum_{J\in\overline{\mathfrak N}_{>I}} {\mathsf H}_{IJ}.
\end{align}
By using the scattering equations \eqref{SE-MRK} and their unique solution \eqref{MRK-sol}, it is easy to obtain
\begin{align}\label{}
  {\mathsf H}_{22} \,&=\, 1 - \sum_{I\in\overline{\mathfrak N}} \tau_I \,=\, \prod_{I\in\overline{\mathfrak N}} {q_I^\perp \over q_{I+1}^\perp},
  \\
  {\mathsf H}_{II} \,&=\, \zeta_I \bigg(1  -  \sum_{J\in\overline{\mathfrak N}_{>I}} {\tau_J} \bigg)
  - \tau_I  \bigg(1 + \sum_{J\in\overline{\mathfrak N}_{<I}} {\zeta_J} \bigg)
  \,=\,  x_I \zeta_I\tau_I\, (v_I^\ast  + x_I^\ast).
\end{align}

Then we have
\begin{align}\label{det-H-neg-x1}
   {\det}'\, {\mathsf H} \,=\, 
   \Bigg(\prod_{I\in\overline{\mathfrak N}} x_I \zeta_I \tau_I \Bigg)\, \det {\sf H}',
\end{align}
where one choose to remove the first column and row corresponding to particle label `1', and ${\mathsf H}'$ is defined as
\begin{align}\label{}
   {\mathsf H}' \,=\, 
   \left(\begin{matrix}
       {\mathsf H}_{22} & c_{2 I_1} & c_{2 I_2}  & \cdots & c_{2 I_m} 
       \\
       x_{I_1}^\ast & v_{I_1}^\ast {+} x_{I_1}^\ast & c_{I_1 I_2} x_{I_1}^\ast  & \cdots & c_{I_1 I_m}x_{I_1}^\ast
       \\
       x_{I_2}^\ast & x_{I_2}^\ast & v_{I_2}^\ast {+} x_{I_2}^\ast  & \cdots & c_{I_2 I_m}x_{I_2}^\ast
       \\
       \vdots & \vdots & \vdots & \ddots & \vdots 
       \\
       x_{I_m}^\ast & x_{I_m}^\ast & x_{I_m}^\ast  & \cdots & v_{I_m}^\ast {+} x_{I_m}^\ast
       \\
   \end{matrix}\right),
\end{align}
where $I_1 < \cdots < I_m \in\overline{\mathfrak N}$, $m = k{-}2$.
This matrix again displays a similar structure as $\bar\phi$ in \eqref{MHV-phi-matrix-MRK}.
Hence we can calculate its determinant by the similar technique based on the matrix determinant lemma.
The final result is
\begin{align}\label{theorem-3}
\boxed{
  {\det}'\,{\mathsf H} \,=\, \Bigg(\prod_{I\in\overline{\mathfrak N}} x_I \zeta_I\tau_I \Bigg)
  \prod_{I\in\overline{\mathfrak N}} v_I^\ast.
}
\end{align}
The details of the derivation can be found in Appendix \ref{app-determinents}.


Putting everything together, we obtain that any N$^{k}$MHV amplitude of graviton factorizes as
\begin{align}\label{}
  {\cal M}_n \,&\simeq\, -s^2\, {\cal C}(2;3) {-1 \over |{q_{4}^\perp}|^2} {\cal V}(q_{4}; 4; q_{5}) \cdots
  {-1 \over |{q_{n-1}^\perp}|^2} {\cal V}(q_{n-1}; n{-}1; q_{n}) {-1 \over |{q_{n}^\perp}|^2}\, {\cal C}(1; n)\,,
\end{align}
with
\begin{align}
  {\cal V}(q_i, {i^+}, q_{i+1})
  \,&=\, q_i^{\perp^\ast} v_i\, q_{i+1}^{\perp}
  \,=\, {q_i^{\perp^\ast} \big( k_{i}^{\perp} q_{i}^{\perp^\ast}  - k_{i}^{\perp^\ast} q_{i}^{\perp} \big) q_{i+1}^{\perp} \over (k_i^\perp)^2},
  \\
  {\cal V}(q_I, {I^-}, q_{I+1}) \,&=\,  q_I^{\perp} v_{I}^\ast\, q_{I+1}^{\perp^\ast}
  \,=\, {q_I^{\perp} \big( k_I^{\perp^\ast} q_I^{\perp}  -  k_I^{\perp} q_I^{\perp^\ast} \big) q_{I+1}^{\perp^\ast} \over (k_I^{\perp^\ast})^2},
\end{align}
in agreement with Lipatov formula \eqref{MRK-GR-main}.

Let us conclude this section by making some comments.
First, we have assumed in previous sections that the two incoming particles 1 and 2 carry the same helicity.
Here we show that all conclusions hold for the case where the gravitons 1 and 2 have opposite helicities.
For example, let us consider amplitude ${\cal M}_n$ with helicity configuration $(1^+,2^-,\ldots,n^-)$.
In this case, in MRK, we have
\begin{align}
  {\cal M}_n(1^+,2^-,\ldots,n^-) \,=\, {\cal M}_n(1^-,2^-,\ldots, n^+) \left({1 \over (1\,n)^8}\bigg|_\text{MRK}\right).
\end{align}
The factor $(1\,n)^{-8}$ combines with the impact factor ${\cal C}(1^+; n^-)$ to give
\begin{align}
  \left({1 \over (1\,n)^8}\bigg|_\text{MRK}\right)
  {\cal C}(1^-; n^+) \,=\,  {\cal C}(1^-; n^+)^\ast \,=\, {\cal C}(1^+; n^-).
\end{align}
Second, we would like show that the helicity is conserved in gravitational impact factors, like gauge theory.
As an example, we consider amplitude ${\cal M}_n(1^-,2^-,3^-,4^+,\ldots)$.
In the multi-Regge limit, we have
\begin{align}
  {\cal M}_n(1^-,2^-,3^-,4^+,\ldots) \,&=\, {\cal M}_n(1^-,2^-,3^+,4^-,\ldots) 
  \left({1 \over (3\,4)^8}\bigg|_\text{MRK}\right)
  \nonumber\\
  &=\, {\cal M}_n(1^-,2^-,3^+,4^+,\ldots) 
  \left(-{(k_4^\perp)^2\, q_4^{\perp} q_{5}^{\perp^\ast}  \over (k_4^{\perp^\ast})^2\, q_4^{\perp^\ast} q_{5}^{\perp}}\right)
  \bigg({1 \over (3\,4)^8}\bigg|_\text{MRK}\bigg).
\end{align}
This shows that the amplitude ${\cal M}_n(1^-,2^-,3^-,4^+,\ldots)$ is suppressed in MRK since
\begin{align}
  \bigg({1 \over (3\,4)^8}\bigg|_\text{MRK}\bigg) \,\simeq\, {\cal O}\Big( \big(k_4^+/ k_3^+\big)^4\Big).
\end{align}
Finally, let us see what happens in the case where particles with other spins in supergravity are involved.
For example, in case of one pair of gravitinos, e.g.~$(1_{\tilde G}^-, n_{\tilde G}^+)$
\begin{align}
  {\cal M}_n(1_{\tilde G}^-,n_{\tilde G}^+, \ldots) \,=\, {\cal M}_n(1^-,n^+,\ldots) \left({-1 \over (1\,n)}\bigg|_\text{MRK}\right).
\end{align}
Similarly, a combination of the factor $(1\,n)^{-1}$ and the impact factor ${\cal C}(1^+; n^-)$ gives
\begin{align}
  \left({-1 \over (1\,n)}\bigg|_\text{MRK}\right)
  {\cal C}(1^-; n^+) \,=\,  
  \left({k_n^{\perp^\ast} \over k_n^\perp}\right)^{3/2}
  \,=\, {\cal C}(1_{\tilde G}^-; n_{\tilde G}^+),
\end{align}
which exactly agrees with the result in \cite{Lipatov:1982it,Lipatov:1982vv}.
It is also easy to obtain the result for other cases where particles with other spins in the supergravity multiplet are involved in the same way.

\section{Conclusions}

We have initiated the investigation of the gravitational scattering in the multi-Regge regime in the framework of the scattering equations.
Unlike gauge theory, the evaluation of the determinants of the two matrices is involved in the formula of graviton amplitudes. 
In general, it seems impossible to obtain the exact compact results of these determinants even in the MHV sector.
We have shown that the two matrices get greatly simplified in the multi-Regge limit, and finally we can obtain compact expressions for any N$^k$MHV sector for any number of external particles.
As a consequence, we provide an elegant derivation of the tree-level multi-Regge factorization of gravitational scattering amplitudes.

It should be emphasized that our analysis in this paper is based on the asymptotic behaviour of the solutions to the scattering equations in the multi-Regge limit.
In \cite{Duhr:2018ppq}, it is conjectured that all solutions of the scattering equations admit the same hierarchy as the rapidity ordering in MRK, as shown in \eqref{conjecture} in this paper.
While we do currently not have a rigorous mathematical proof, the result in this paper provides very strong support to the validity of the conjecture.
As a next step, it would also be interesting to investigate the multi-Regge limit of amplitudes for more theories along this path.

While this paper has been concentrated on the multi-Regge limit, it would be interesting to study graviton amplitudes in various generalizations of MRK, where two or more produced particles have comparable rapidities.
We leave this study for future work.

\section*{Acknowledgements}

I would especially like to thank Xiaoran Zhao for stimulating discussions, and Claude Duhr for useful discussions and a careful reading of the manuscript, as well as collaboration on related topics.
I am also grateful to Song He and Brenda Penante for discussions.
I would like to acknowledge the hospitality of CERN Theory Division in Geneva, Galileo Galilei Institute in Firenze, ITP, CAS in Beijing, and Tianjin University in Tianjin, where part of the work was done.
This work was supported by the ``Fonds Sp\'{e}cial de Recherche'' (FSR) of the UCLouvain.

\appendix
\section{Matrix determinant lemma}\label{app-mdl}
Let $\mathbf{A}$ be an invertible matrix, $\mathbf{u}$ and $\mathbf{v}$ be two column vectors. Then
the matrix determinant lemma states that \cite{Harville-1997} (c.f.~also \cite{Ding-Zhou-MDL-2007})
\begin{align}\label{MDL} 
    \det\big(\mathbf{A}+\mathbf{u}\mathbf{v}^\mathrm{T}\big) \,=\, \big(1 + \mathbf{v}^\mathrm{T}\mathbf{A}^{-1}\mathbf{u}\big)\,\det(\mathbf{A}).
\end{align}

\noindent{\it Proof:~}
Let us first see the special case of the identity matrix, i.e.\,$\mathbf{A}=\mathds{1}$. Note
\begin{align}\label{} 
   \left(\begin{matrix}
       \mathds{1} & \mathbf{0}   \\
       \mathbf{v}^\mathrm{T} & 1 \\
  \end{matrix}\right)
  \left(\begin{matrix}
       \mathds{1}+\mathbf{u}\mathbf{v}^\mathrm{T} & \mathbf{u}   \\
       \mathbf{0} & 1 \\
  \end{matrix}\right)
  \left(\begin{matrix}
       \mathds{1} & \mathbf{0}   \\
       -\mathbf{v}^\mathrm{T} & 1 \\
  \end{matrix}\right)
  \,=\,
  \left(\begin{matrix}
       \mathds{1} & \mathbf{u}   \\
       \mathbf{0} & 1 + \mathbf{v}^\mathrm{T}\mathbf{u} \\
  \end{matrix}\right).
\end{align}
This ends the proof for the case of $\mathbf{A}=\mathds{1}$.
Then from
\begin{align}\label{} 
   \mathbf{A} + \mathbf{u}\mathbf{v}^\mathrm{T} \,=\, \mathbf{A} \big( \mathds{1} + \mathbf{A}^{-1}\mathbf{u}\mathbf{v}^\mathrm{T} \big),
\end{align}
we can prove the lemme, i.e.,
\begin{align}\label{} 
    \det\big(\mathbf{A}+\mathbf{u}\mathbf{v}^\mathrm{T}\big) 
    \,=\, \det\big(\mathbf{A}\big) \det\big( \mathds{1} + \mathbf{A}^{-1}\mathbf{u}\mathbf{v}^\mathrm{T} \big)
    \,=\, \det(\mathbf{A})\, \big(1 + \mathbf{v}^\mathrm{T}\mathbf{A}^{-1}\mathbf{u}\big).
\end{align}

\section{Proof of three identities}\label{app-determinents}
This appendix provides the details of deriving three identities used in section \ref{GR-MRK}, i.e.~eqs.\,\eqref{mhv-mdl-result}, \eqref{theorem-2} and \eqref{theorem-3}.

\subsection{Equation \eqref{mhv-mdl-result}}

In this section, we focus on the following triangular matrix:
\begin{align}\label{}
   \varphi \,=\, 
   \left(\begin{matrix}
       v_4 & x_5 {-} x_4 {-} v_4 & \cdots & x_{n-1} {-} x_4 {-} v_4 & x_n {-} x_4   \\
       0 & v_5 & \cdots & x_{n-1} {-} x_5 & x_n {-} x_5   \\
       \vdots & \vdots & \ddots & \vdots & \vdots \\
       0 &  0 & \cdots & v_{n-1} & x_n {-} x_{n-1} \\
       0 & 0 & \cdots & 0 & x_n \\
  \end{matrix}\right).
\end{align}
Our goal is to find $(\varphi^{-1})_{1,n-3}$.
Let us denote the last column of the inverse of this matrix as $\alpha=(\alpha_{4}, \alpha_{5}, \ldots, \alpha_n)^\mathrm{T}$, and then it satisfies the following equation
\begin{align}\label{}
   \varphi\,\alpha \,=\, (0, 0, \ldots, 1)^\mathrm{T}.
\end{align}
In the following we show that its solution is
\begin{align}
  \label{mhv-alpha-i}
   \alpha_{i} \,&=\, {k_i^\perp \over k_n^{\perp^\ast}}, \quad i=n, n{-}1, \ldots, 5  \\
   \alpha_{4} \,&=\, -{k_3^\perp + k_n^\perp \over k_n^{\perp^\ast}}.
   \label{mhv-alpha-4}
\end{align}
First, we can easily get from $x_n \alpha_n=1$
\begin{align}\label{}
   \alpha_{n} \,&=\, {1 \over x_n} \,=\, {k_n^\perp \over k_n^{\perp^\ast}}.
\end{align}
Then we assume all $\alpha_j$ with $j>i$ are given by \eqref{mhv-alpha-i}, and let us solve $\alpha_i$ from the following equation:
\begin{align}\label{}
   v_i \alpha_i + \sum_{j=i+1}^n (x_j-x_i)\alpha_j \,=\, 0.
\end{align}
Plugging the values of $\alpha_j$ ($j>i$) given in \eqref{mhv-alpha-i} into this equation gives
\begin{align}\label{}
  \alpha_i \,=\, - {1 \over v_i} \sum_{j=i+1}^n (x_j-x_i)\alpha_j \,=\, {k_i^\perp \over k_n^{\perp^\ast}},
\end{align}
which agrees with eq.\,\eqref{mhv-alpha-i}. Finally, solving the last equation gives eq.\,\eqref{mhv-mdl-result}, i.e.,
\begin{align}\label{}
  (\varphi^{-1})_{1,n-3} \,\equiv\, \alpha_4 \,=\,  - {1 \over v_4} 
  \Bigg(\sum_{j=5}^n (x_j-x_4)\alpha_j - v_4 \sum_{j=5}^{n-1} \alpha_j \Bigg) \,=\, -{k_3^\perp + k_n^\perp \over k_n^{\perp^\ast}}.
\end{align}

\subsection{Equation \eqref{theorem-2}}
In the following we consider the reduced determinant of the $\overline{\mathsf H}_{k\times k}$ whose indices take values from the set $\mathfrak{P}$.

First fo all, it is useful to introduce a new notation related to particle labels: $I_{\ell_i} \in {\mathfrak N}$ denotes the smallest number that satisfies $I_{\ell_i}>i\in \mathfrak{P}$. For example, $\ell_3=1$ because $3<I_1\in {\mathfrak N}$.
Then, by abuse of multiple subscripts, we can rewrite $\zeta_i$ and $\tau_i$ in terms of  $\zeta_I$ and $\tau_I$ as follows:
\begin{align}\label{sol-i-app}
\begin{aligned}
  \tau_i \,&=\, 
  \begin{cases}
    -{k_{I_{\ell_i}}^{\perp^\ast} \over q^{\perp^\ast}_{I_{\ell_i}}}\, \zeta_{I_{\ell_i}}^{-1}, & \text{if}~i < I_m, \\
     - \prod\limits_{1 \le l \le m} {q_{I_l}^{\perp^\ast}  \over  q_{I_l+1}^{\perp^\ast}}, & \text{if}~i>I_m,
  \end{cases}
\end{aligned}
\qquad
\begin{aligned}
  \zeta_i \,=\, 
  \begin{cases}
      -{k_{I_{\ell_i}}^\perp \over q^\perp_{I_{\ell_i}} }\, \tau_{I_{\ell_i}}^{-1}, ~&\text{if}~i < I_m, \\
      - 1,~&\text{if}~i > I_m.
  \end{cases}
\end{aligned}
\end{align}

Let us also rewrite the matrix \eqref{H-matrix-mrk-pos}:
\begin{align}\label{}
   \overline{\sf H}'  \,&=\, 
   \left(\begin{matrix}
       v_{i_1} {+} x_{i_1} & c_{i_1 i_2} & \cdots & c_{i_1 i_p}  & c_{i_1 n}
       \\
       x_{i_2} & v_{i_2} {+} x_{i_2} & \cdots & c_{i_2 i_p} & c_{i_2 n}
       \\
       \vdots & \vdots & \ddots & \vdots & \vdots 
       \\
       x_{i_p} & x_{i_p} & \cdots & v_{i_p} {+} x_{i_p}  & c_{i_p n}
       \\
       x_{n} & x_{n} & \cdots & x_{n}  & x_{n}
       \\
   \end{matrix}\right),
\end{align}
where particle labels in ${\mathfrak P}$ have been reordered as $3 < i_1 < \cdots < i_p < n$ with $p=(n{-}k) {-} 2$.
By performing some elementary row and column transformations, we have
\begin{align}\label{}
   \det \overline{\sf H}'  \,&=\, \det \overline{\sf H}'',
   \quad
   \overline{\sf H}'' \,=\,
   \begin{pmatrix}
       v_{i_1} & c_{i_1 i_2} {-} x_{i_1} {-} v_{i_1} & \cdots & c_{i_1 i_p} {-} x_{i_1} {-} v_{i_1} & c_{i_1 n} {-} x_{i_1}
       \\
       0 & v_{i_2} & \cdots & c_{i_2 i_p} {-} x_{i_2} & c_{i_2 n} {-} x_{i_2}
       \\
       \vdots & \vdots & \ddots & \vdots  & \vdots
       \\
       0 & 0 & \cdots & v_{i_p}  &  c_{i_p n} {-} x_{i_p}
       \\
       x_{n} & 0 & \cdots &  0  & x_{n}
       \\
   \end{pmatrix}.
\end{align}
Using the matrix determinant lemma, we have
\begin{align}\label{MRK-H-pos-MRL}
   \det \overline{\mathsf H}' \,&=\,  x_n\, \Bigg(\prod_{i\in {\mathfrak P}, i\ne 3,n} v_i \Bigg)\,
   \Big( 1 + x_n \left(\bar{\sf \Phi}^{-1}\right)_{1, p+1} \Big),
\end{align}
where $\bar{\sf \Phi}$ is nothing but $\overline{\sf H}''$ with replacing the first element of the last row $x_n$ by zero.
Now our task becomes to calculate the entry in the upper right corner of the inverse of the matrix $\bar{\sf \Phi}$.
Let us denote the last column of the inverse of $\bar{\sf \Phi}$ as:
\begin{align}\label{}
   (\alpha_{i_1}, \alpha_{i_2}, \ldots, \alpha_{i_p}, \alpha_n)^\mathrm{T}.
\end{align}
Clearly, $\alpha_{i_1}=(\bar{\sf \Phi}^{-1})_{1, p+1}$.
Then it can be determined by the following linear equations:
\begin{align}\label{eq-linear-pos}
   \bar{\sf \Phi}\, (\alpha_{i_1}, \alpha_{i_2}, \ldots, \alpha_n)^\mathrm{T} \,=\, (0, 0, \ldots,1)^\mathrm{T}.
\end{align}
The solution is
\begin{align}
  \alpha_j \,&=\,  -{k_j^\perp \over k_n^{\perp^\ast}}\,\zeta_j, \quad i_1 < j \le n,
  \label{MRK-det-Hp-mathInduction-1}
  \\
  \alpha_{i_1} \,&=\, x_n^\ast \,\left( {k_3^\perp \over k_n^\perp}\zeta_3 - 1\right).
  \label{MRK-det-Hp-mathInduction-2}
\end{align}
Plugging \eqref{MRK-det-Hp-mathInduction-1} into \eqref{MRK-H-pos-MRL} immediately leads to
\begin{align}\label{}
   \det \overline{\mathsf H}' \,&=\,  
   - x_n\, \Bigg(\prod_{i\in {\mathfrak P}, i\ne 3,n} v_i \Bigg)\,{k_3^\perp \over k_n^\perp}\, \zeta_3\,\tau_3\,.
\end{align}
We immediately obtain \eqref{theorem-2} by inserting $\det \overline{\mathsf H}'$ into \eqref{det-H-pos-x1}.

We show how to obtain \eqref{MRK-det-Hp-mathInduction-1} and \eqref{MRK-det-Hp-mathInduction-2} by induction in the following.
First, it is very easy to obtain $\alpha_n$ from equation \eqref{eq-linear-pos},
\begin{align}\label{}
  \alpha_n \,=\, x_n^\ast \,=\,  -{k_n^\perp \over k_n^{\perp^\ast}}\,  \zeta_n,
\end{align}
where one uses the solution of the MRK scattering equations, $\zeta_n=-1$.
As a next step, we assume all $\alpha_j$'s ($j>i$) are given by \eqref{MRK-det-Hp-mathInduction-1}, and then let us solve $\alpha_i$ which satisfies the following equation:
\begin{align}\label{det-H-pos-DML-X-eq}
  v_i \alpha_i + \sum_{j \in {\mathfrak P}, j > i} \big( c_{ij} - x_i \big) \alpha_j \,=\, 0, \qquad i>i_1.
\end{align}
Using the definition of $c_{ij}$ and $\alpha_j$ given by \eqref{MRK-det-Hp-mathInduction-1}, we have
\begin{align}\label{}
  \sum_{j \in {\mathfrak P}, j > i} \big( c_{ij} - x_i \big) \alpha_j   \,&=\,
  \sum_{j \in {\mathfrak P}, j > i} \Bigg( - {\zeta_i \tau_j  \over  \tau_i} {k_j^{\perp^\ast} \over k_n^{\perp^\ast}} 
  + x_i {k_j^\perp \over k_n^{\perp^\ast}}\,\zeta_j\Bigg)
  \nonumber\\
  \,&=\,
  {\zeta_i \over k_n^{\perp^\ast}} \sum_{j \in {\mathfrak P}, j > i} \Bigg( - {\tau_j k_j^{\perp^\ast} \over \tau_i} 
  + x_i {\zeta_j k_j^\perp \over \zeta_i} \Bigg)
  \nonumber\\
  \,&=\,
  {\zeta_i \over k_n^{\perp^\ast}} 
  \Bigg( - {{\cal T}_i \over \tau_i} +  {x_i {\cal Z}_i \over \zeta_i} \Bigg),
  \label{th2-app-eq-x1}
\end{align}
where we denote
\begin{align}\label{}
  {\cal T}_i
  \,&=\,
  \sum_{j \in {\mathfrak P}, j > i}  \tau_j k_j^{\perp^\ast},
  \qquad
  {\cal Z}_i  \,=\, \sum_{j \in {\mathfrak P}, j > i} {\zeta_j k_j^\perp}.
\end{align}
Next, we calculate these tow terms for two cases respectively:~the label $i$ is bigger than the label of any negative-helicity particle or not.
\begin{itemize}
  \item Let us first consider the case of the label $i$ is less than the largest label carried by negative-helicity particles, i.e.\,$i<I_m$. In this case, we have
  \begin{align}
       {\cal T}_i   \,&=\,
       \sum_{j \in {\mathfrak P}, i< j < I_{\ell_i}}  \tau_j k_j^{\perp^\ast} \,+\, \sum_{j \in {\mathfrak P}, j > I_{\ell_i}}  \tau_j k_j^{\perp^\ast}
       \\
       \,&=\,  \tau_i\,\sum_{i< j < I_{\ell_i}} k_j^{\perp^\ast} \,+\, {k_{I_{\ell_i}}^{\perp^\ast} \over \zeta_{I_{\ell_i}}}
       \label{Ti-coe-app-2}
       \\
       \,&=\,  \tau_i\,\sum_{i< j < I_{\ell_i}} k_j^{\perp^\ast} \,-\, \tau_i\, q_{I_{\ell_i}}^{\perp^\ast}
       \label{Ti-coe-app-3}
       \\
       \,&=\,  - \tau_i\, q_{i+1}^{\perp^\ast}.
  \end{align}
  Here we used the scattering equations $\bar{\cal S}_I^{\dot 2}=0$, \eqref{SE-MRK}, in the second line \eqref{Ti-coe-app-2}, and the solution \eqref{sol-i-app} in the third line \eqref{Ti-coe-app-3}.
  Similarly, for ${\cal Z}_i$ we have
  \begin{align}\label{}
      {\cal Z}_i  \,&=\, - \sum_{j \in {\mathfrak P}, j \le i} {\zeta_j k_j^\perp}
      \\
      \,&=\, - \sum_{j \in {\mathfrak P}, j < I_{\ell_i}} {\zeta_j k_j^\perp} \,+\, \sum_{j \in {\mathfrak P}, i< j < I_{\ell_i}} {\zeta_j k_j^\perp}
      \\
      \,&=\, {k_{I_{\ell_i}}^\perp \over \tau_{I_{\ell_i}}} \,+\, \zeta_i \sum_{j \in {\mathfrak P}, i< j < I_{\ell_i}} k_j^\perp
      \\
      \,&=\, - \zeta_i\, q_{I_{\ell_i}}^\perp \,+\, \zeta_i \sum_{j \in {\mathfrak P}, i< j < I_{\ell_i}} k_j^\perp
      \\
      \,&=\, - \zeta_i\, q_{i+1}^\perp
  \end{align}
  
  \item In the other case, i.e.\,$i>I_m$, it is easy to obtain
  \begin{align}\label{}
      {\cal T}_i  \,&=\,  \tau_i \sum_{i < j \le n}  k_j^{\perp^\ast} \,=\,  - \tau_i\, q_{i+1}^{\perp^\ast},
      \\
      {\cal Z}_i  \,&=\, - \zeta_i \sum_{i < j \le n} {k_j^\perp} \,=\, - \zeta_i\, q_{i+1}^\perp.
  \end{align}
\end{itemize}
In both cases, as expected, we obtain the same results for ${\cal T}_i$ and ${\cal Z}_i$.
By inserting them into \eqref{th2-app-eq-x1}, we find
\begin{align}\label{}
  \sum_{j \in {\mathfrak P}, j > i} \big( c_{ij} - x_i \big) \alpha_j   \,&=\,
  {\zeta_i \over k_n^{\perp^\ast}} 
  \Bigg( - {{\cal T}_i \over \tau_i} +  {x_i {\cal Z}_i \over \zeta_i} \Bigg)
  \,=\,  {k_i^\perp \over k_n^{\perp^\ast}}\,v_i\, \zeta_i\,.
\end{align}
Finally, equation\,\eqref{det-H-pos-DML-X-eq} can be solved exactly by
\begin{align}\label{}
  \alpha_i  \,=\,  - {k_i^\perp \over k_n^{\perp^\ast}}\,\zeta_i,
\end{align}
which proves \eqref{MRK-det-Hp-mathInduction-1}.

As a final step, we prove \eqref{MRK-det-Hp-mathInduction-2} via finding $\alpha_{i_1}$ which satisfies
\begin{align}\label{det-H-pos-DML-X-eq-1}
  v_{i_1} \alpha_{i_1} + \sum_{j \in {\mathfrak P}, j > i_1} \big( c_{i_1 j} - x_{i_1} - v_{i_1} \big) \alpha_j + v_{i_1}\alpha_n \,=\, 0.
\end{align}
Noting that
\begin{align}\label{}
  \sum_{j \in {\mathfrak P}, j > i_1} \big( c_{i_1 j} - x_{i_1} - v_{i_1} \big) \alpha_j
  \,&=\,
  {\zeta_{i_1} \over k_n^{\perp^\ast}}  \Bigg( - {{\cal T}_{i_1} \over \tau_{i_1}} +  (x_{i_1} + v_{i_1}){{\cal Z}_{i_1} \over \zeta_{i_1}} \Bigg)
  \,=\,  - {q_{i_1}^\perp \over k_n^{\perp^\ast}}\,v_{i_1}\, \zeta_{i_1},
\end{align}
we have
\begin{align}\label{}
  \alpha_{i_1}  \,&=\,  {q_{i_1}^\perp \over k_n^{\perp^\ast}}\, \zeta_{i_1} - x_n^\ast
  \,=\, - x_n^\ast \bigg( {k_3^\perp \over k_n^\perp}\, \zeta_3\,\tau_3 + 1\bigg),
\end{align}
which ends the proof.

\subsection{Equation \eqref{theorem-3}}
Let us now discuss another part corresponding to the set $\mathfrak{N}$.
Our goal is to evaluate the determinant of the following matrix:
\begin{align}\label{}
   {\mathsf H}' \,=\, 
   \left(\begin{matrix}
       {\mathsf H}_{22} & c_{2 I_1} & c_{2 I_2}  & \cdots & c_{2 I_m} 
       \\
       x_{I_1}^\ast & v_{I_1}^\ast {+} x_{I_1}^\ast & c_{I_1 I_2} x_{I_1}^\ast  & \cdots & c_{I_1 I_m}x_{I_1}^\ast
       \\
       x_{I_2}^\ast & x_{I_2}^\ast & v_{I_2}^\ast {+} x_{I_2}^\ast  & \cdots & c_{I_2 I_m}x_{I_2}^\ast
       \\
       \vdots & \vdots & \vdots & \ddots & \vdots 
       \\
       x_{I_m}^\ast & x_{I_m}^\ast & x_{I_m}^\ast  & \cdots & v_{I_m}^\ast {+} x_{I_m}^\ast
       \\
   \end{matrix}\right)
\end{align}
with
\begin{align}\label{}
  {\mathsf H}_{22} \,=\,  \prod_{I\in\overline{\mathfrak N}} {q_I^\perp \over q_{I+1}^\perp},
  \qquad
  c_{2I} \,=\, {\tau_I \over \zeta_I},
  \qquad
  c_{IJ} \,=\, {\zeta_I \tau_J \over  \tau_I \zeta_J}~~\text{for}~~I<J.
\end{align}

Using a little linear algebra, it is easy to find
\begin{align}\label{}
  \det {\mathsf H}' \,=\,
   \left|\begin{matrix}
       {\mathsf H}_{22} & c_{2 I_1} {-} {\sf H}_{22} & c_{2 I_2} {-} {\sf H}_{22}  & \cdots & c_{2 I_m} {-} {\sf H}_{22}
       \\
       0 & v_{I_1}^\ast & (c_{I_1 I_2} {-} 1) x_{I_1}^\ast  & \cdots & (c_{I_1 I_m} {-} 1) x_{I_1}^\ast - x_{I_1}^\ast x_{I_m} v_{I_m}^\ast
       \\
       0 & 0 & v_{I_2}^\ast  & \cdots & (c_{I_2 I_m} {-} 1) x_{I_2}^\ast - x_{I_2}^\ast x_{I_m} v_{I_m}^\ast
       \\
       \vdots & \vdots & \vdots & \ddots & \vdots 
       \\
       x_{I_m}^\ast & 0 & 0  & \cdots & v_{I_m}^\ast
       \\
   \end{matrix}\right|.
\end{align}
By the matrix determinant lemma, we have
\begin{align}\label{det-H-neg-MDL-main-X}
   \det {\mathsf H}' \,&=\,  {\mathsf H}_{22}\, \Bigg(\prod_{I\in\overline{\mathfrak N}} v_I^\ast \Bigg)\,
   \Big( 1 + x_{I_m}^\ast \big({\sf \Phi}^{-1}\big)_{1, m+1} \Big),
\end{align}
where
\begin{align}\label{}
   {\sf \Phi}  \,&=\, 
   \left(\begin{matrix}
       {\mathsf H}_{22} & c_{2 I_1} {-} {\sf H}_{22} & c_{2 I_2} {-} {\sf H}_{22}  & \cdots & c_{2 I_m} {-} {\sf H}_{22}
       \\
       0 & v_{I_1}^\ast & (c_{I_1 I_2} {-} 1) x_{I_1}^\ast  & \cdots & (c_{I_1 I_m} {-} 1) x_{I_1}^\ast - x_{I_1}^\ast x_{I_m} v_{I_m}^\ast
       \\
       0 & 0 & v_{I_2}^\ast  & \cdots & (c_{I_2 I_m} {-} 1) x_{I_2}^\ast - x_{I_2}^\ast x_{I_m} v_{I_m}^\ast
       \\
       \vdots & \vdots & \vdots & \ddots & \vdots 
       \\
       0 & 0 & 0  & \cdots & v_{I_m}^\ast
       \\
   \end{matrix}\right).
\end{align}
Thus now our task is to calculate the last entry in the first row of the inverse of this matrix.
As we will see in the following, the similar technique used in previous sections also works in this case.

Let us denote the first row of the inverse of the ${\sf \Phi}$ as:
\begin{align}\label{}
   (\bar\alpha_0, \bar\alpha_1, \ldots, \bar\alpha_m)
\end{align}
They can be determined by the following linear equations:
\begin{align}\label{det-H-neg-eq-main-matrix}
   (\bar\alpha_0, \bar\alpha_1, \ldots, \bar\alpha_m)\, {\sf \Phi}  \,=\, (1, 0, \ldots, 0)
\end{align}
The solution of this equation is
\begin{align}
  \bar\alpha_0 \,&=\, {1 \over {\sf H}_{22}} \,=\, \prod_{I\in\overline{\mathfrak N}} {q_{I+1}^\perp \over q_I^\perp},
  \label{MRK-det-Hn-mathInduction-0}
  \\
  \bar\alpha_a  \,&=\, \bar\alpha_0\, x_{I_a} \tau_{I_a}~~~\text{for}~~ 1 \le a < m,
  \label{MRK-det-Hn-mathInduction-I}
  \\
  \bar\alpha_m  \,&=\, \big(\bar\alpha_0 - 1\big) x_{I_m}.
  \label{MRK-det-Hn-mathInduction-m}
\end{align}
In the following, we prove them by induction.
First of all, it is very easy to obtain $\bar\alpha_0$ and $\bar\alpha_1$ by solving the first two equations in \eqref{det-H-neg-eq-main-matrix}.
In next step, we assume all $\bar\alpha_b$'s ($b<a$) are given by \eqref{MRK-det-Hn-mathInduction-0} and \eqref{MRK-det-Hn-mathInduction-I}, then let us solve $\bar\alpha_a$ which satisfies the following equation:
\begin{align}\label{det-H-neg-DML-X-eq-main}
  v_{I_a}^\ast \bar\alpha_a
  + \big(c_{2I_a} - {\sf H}_{22} \big)\bar\alpha_0 
  + \sum_{r=1}^{a-1}\big(c_{I_r I_a} - 1\big)x_{I_r}^\ast \bar\alpha_r
  \,=\, 0
  \quad\text{for}~~1 < a < m.
\end{align}
Here we first consider the second term on the left-hand side of the equation.
By observing the solution of the MRK scattering equations, \eqref{MRK-sol}, we find that ${\sf H}_{22}$ and $c_{2I}$ can be written as
\begin{align}\label{}
  {\mathsf H}_{22} \,&=\,  \prod_{I\in\overline{\mathfrak N}} {q_I^\perp \over q_{I+1}^\perp}
  \,=\, {q_{I_a}^\perp \over k_{I_a}^\perp} \left( \prod_{l=1}^{a-1} {q_{I_l}^\perp \over q_{I_l+1}^\perp} \right) \tau_{I_a},
  \\
  c_{2I_a} \,&=\, {\tau_{I_a} \over \zeta_{I_a}}
  \,=\, {q_{I_a}^{\perp\ast} \over k_{I_a}^{\perp\ast}} \left( \prod_{l=1}^{a-1} {q_{I_l}^{\perp\ast} \over q_{I_l+1}^{\perp\ast}} \right) \tau_{I_a}.
\end{align}
Therefore we have
\begin{align}\label{}
  \big(c_{2I_a} - {\sf H}_{22} \big)\bar\alpha_0 
  \,=\, \big(h_a - h_a' \big) \bar\alpha_0\,x_{I_a} \tau_{I_a},
\end{align}
where
\begin{align}\label{}
  h_a \,\equiv\, {k_{I_a}^{\perp} q_{I_a}^{\perp^\ast} \over 
  \left(k_{I_a}^{\perp^\ast}\right)^2} \left( \prod_{l=1}^{a-1} {q_{I_l}^{\perp^\ast} \over q_{I_l+1}^{\perp^\ast}} \right),
  \qquad
  h_a' \,\equiv\, {q_{I_a}^\perp \over k_{I_a}^{\perp^\ast}} \left( \prod_{l=1}^{a-1} {q_{I_l}^\perp \over q_{I_l+1}^\perp} \right).
\end{align}

Let us now turn to the last term on the left-hand side of \eqref{det-H-neg-DML-X-eq-main}.
Comparing to the solution of the MRK scattering equations \eqref{MRK-sol}, we have
\begin{align}\label{}
  \tau_{I_r}  \,&=\,  
  {k_{I_r}^\perp q_{I_a+1}^\perp \over k_{I_a}^\perp q_{I_r+1}^\perp} \left( \prod_{l=r+1}^{a} {q_{I_l}^\perp \over q_{I_l+1}^\perp} \right) \tau_{I_a},
  \quad r<a.
\end{align}
This leads to
\begin{align}\label{}
  \big(c_{I_r I_a} - 1\big)x_{I_r}^\ast \bar\alpha_r \,&=\, \big(c_{I_r I_a} - 1\big)x_{I_r}^\ast 
  \bar\alpha_0\, x_{I_r}\,
  {k_{I_r}^\perp q_{I_a+1}^\perp \over k_{I_a}^\perp q_{I_r+1}^\perp} \left( \prod_{l=r+1}^{a} {q_{I_l}^\perp \over q_{I_l+1}^\perp} \right) \tau_{I_a}
  \nonumber\\
  \,&=\, \big(f_{ra} - g_{ra}\big) 
  \big(\bar\alpha_0\, x_{I_a} \tau_{I_a} \big),
\end{align}
where we introduce some short-handed notations:
\begin{align}\label{}
  g_{ra}  \,&=\, {k_{I_r}^\perp q_{I_a+1}^\perp \over k_{I_a}^{\perp^\ast} q_{I_r+1}^\perp} \left( \prod_{l=r+1}^{a} {q_{I_l}^\perp \over q_{I_l+1}^\perp} \right),
  \\
  f_{ra}  \,&=\,  
  {k_{I_r}^\perp q_{I_a+1}^\perp \over k_{I_a}^{\perp^\ast} q_{I_r+1}^\perp} \left( \prod_{l=r+1}^{a} {q_{I_l}^\perp \over q_{I_l+1}^\perp} \right) 
  c_{I_r I_a}
  \,=\,
  {k_{I_a}^{\perp} k_{I_r}^{\perp^\ast}  q_{I_a}^{\perp^\ast}  \over \left(k_{I_a}^{\perp^\ast}\right)^2 q_{I_r}^{\perp^\ast}} 
  \left( \prod_{l=r}^{a-1} {q_{I_l}^{\perp^\ast} \over q_{I_l+1}^{\perp^\ast}} \right).
\end{align}

Therefore, equation \eqref{det-H-neg-DML-X-eq-main} becomes
\begin{align}\label{det-H-neg-DML-X-eq-main-1}
  v_{I_a}^\ast \bar\alpha_a  +  
  \Big[ \big(h_a - h_a' \big)  +  \sum_{r=1}^{a-1} \big(f_{ra} - g_{ra}\big) \Big]
  \big(\bar\alpha_0\, x_{I_a} \tau_{I_a} \big)
  \,=\, 0
  \quad\text{for}~~1 < a < m.
\end{align}
Then by performing a lot of straightforward calculations, we obtain
\begin{align}\label{}
  h_a  +  \sum_{r=1}^{a-1} f_{ra}  \,&=\,  {k_{I_a}^{\perp} q_{I_a}^{\perp^\ast} \over \left(k_{I_a}^{\perp^\ast}\right)^2},
  \\
  h_a' + \sum_{r=1}^{a-1} g_{ra}  \,&=\,  {q_{I_a}^\perp \over k_{I_a}^{\perp^\ast}}.
\end{align}
By plugging them into \eqref{det-H-neg-DML-X-eq-main-1} gives immediately
\begin{align}\label{}
  \bar{\alpha}_a \,=\, \bar\alpha_0\, x_{I_a} \tau_{I_a}.
\end{align}

\vskip 5pt
As a final step, we can obtain $\bar\alpha_m$ by solving the following equation:
\begin{align}\label{det-H-neg-DML-X-eq-final}
  v_{I_m}^\ast \bar\alpha_m
  + \big(c_{2I_m} - {\sf H}_{22} \big)\bar\alpha_0 
  + \sum_{r=1}^{m-1}\big(c_{I_r I_m} - 1\big)x_{I_r}^\ast \bar\alpha_r
  - \sum_{r=1}^{m-1}x_{I_r}^\ast x_{I_m} v_{I_m}^\ast \bar\alpha_r  \,=\, 0.
\end{align}
For the second and the third terms, we have
\begin{align}\label{}
  \big(c_{2I_m} - {\sf H}_{22} \big)\bar\alpha_0 + \sum_{r=1}^{m-1}\big(c_{I_r I_m} - 1\big)x_{I_r}^\ast \bar\alpha_r
  \,=\, - v_{I_m}^\ast \bar\alpha_0\, x_{I_m} \tau_{I_m}.
\end{align}
For the last term, we have
\begin{align}\label{}
  \sum_{r=1}^{m-1}x_{I_r}^\ast x_{I_m} v_{I_m}^\ast \bar\alpha_r  
  \,&=\, \sum_{r=1}^{m-1}x_{I_r}^\ast x_{I_m} v_{I_m}^\ast \bar\alpha_0\, x_{I_r} \tau_{I_r}
  \,=\, v_{I_m}^\ast \bar\alpha_0\, x_{I_m}  \sum_{r=1}^{m-1}  \tau_{I_r}.
\end{align}
Using the scattering equations \eqref{SE-MRK} and their unique solution \eqref{MRK-sol}, we have
\begin{align}\label{}
  \sum_{r=1}^{m-1}x_{I_r}^\ast x_{I_m} v_{I_m}^\ast \bar\alpha_r  
  \,&=\, v_{I_m}^\ast \bar\alpha_0\, x_{I_m}  \Big(1 - {\sf H}_{22} - \tau_{I_m} \Big).
\end{align}
Finally, we find
\begin{align}\label{}
  \bar\alpha_m \,=\, \bar\alpha_0\, x_{I_m}  \big(1 - {\sf H}_{22} \big)
  \,=\,  x_{I_m}  \big(\bar\alpha_0 - 1\big).
\end{align}
Plugging it into \eqref{det-H-neg-MDL-main-X} gives
\begin{align}\label{}
   \det {\mathsf H}' \,&=\,   \prod_{I\in\overline{\mathfrak N}} v_I^\ast\,.
\end{align}

\end{document}